\documentclass[sigconf,nonacm,natbib=false]{acmart}

\usepackage[utf8]{inputenc}
\usepackage[T1]{fontenc}
\usepackage[american]{babel}

\usepackage[draft]{minted}
\usepackage{csquotes} %

\usepackage{titlesec}

\titleformat*{\subsubsection}{\large\bfseries}\titleformat{\subsubsection}
{\normalfont\normalsize\bfseries}{\thesubsubsection}{1em}{}

\usepackage{booktabs}
\usepackage{listings}
\usepackage{enumitem}
\usepackage{amsmath}
\usepackage{subcaption}
\usepackage{tabularx}
\usepackage{makecell}
\usepackage{environ}
\usepackage{balance}

\usepackage[binary-units]{siunitx}
\DeclareSIUnit{\nothing}{\relax}
\DeclareSIUnit{\x}{\times}
\useshorthands*{"}
\defineshorthand{"-}{\babelhyphen{hard}}

\usepackage{graphicx}
\graphicspath{{./images/}}

\PassOptionsToPackage{
	sorting=nyt,
	maxcitenames=2,
	maxbibnames=8,
	style=ACM-Reference-Format
}{biblatex}
\usepackage{biblatex}
\usepackage{xcolor}
\definecolor{ETHa}{RGB}{31,64,122}      %
\definecolor{ETHb}{RGB}{72,90,44}       %
\definecolor{ETHc}{RGB}{18,105,176}     %
\definecolor{ETHd}{RGB}{114,121,28}     %
\definecolor{ETHe}{RGB}{145,5,106}      %
\definecolor{ETHf}{RGB}{111,111,100}    %
\definecolor{ETHg}{RGB}{168,50,45}      %
\definecolor{ETHh}{RGB}{0,122,150}      %
\definecolor{ETHi}{RGB}{149,96,19}      %

\usepackage{listings}

\definecolor{myyellow}{RGB}{247, 204, 51}
\definecolor{myblue}{RGB}{135, 206, 235}
\definecolor{myred}{RGB}{212, 0, 0}
\definecolor{mygreen}{RGB}{133, 187, 101}
\definecolor{mypink}{RGB}{188, 95, 211}

\usepackage{tikz}
\usetikzlibrary{positioning, arrows, matrix, calc}

\DeclareUnicodeCharacter{0301}{ }

\PassOptionsToPackage{pdfpagelabels=false}{hyperref}
\usepackage{hyperref}

\hypersetup{%
	colorlinks=true,%
	urlcolor=ETHg,%
	linkcolor=ETHc,%
	citecolor=ETHd,%
	breaklinks=true,%
	pageanchor=true,%
}

\makeatletter
\DeclareRobustCommand{\varname}[1]{\begingroup\newmcodes@\mathit{#1}\endgroup}
\makeatother

\newcommand\vldbdoi{XX.XX/XXX.XX}
\newcommand\vldbpages{XXX-XXX}
\newcommand\vldbvolume{15}
\newcommand\vldbissue{1}
\newcommand\vldbyear{2022}
\newcommand\vldbauthors{\authors}
\newcommand\vldbtitle{\shorttitle} 
\newcommand\vldbavailabilityurl{https://github.com/RumbleDB/rumbleml-experiments}
\newcommand\vldbpagestyle{plain} 

\begin{document}

\title{RumbleML: program the lakehouse with JSONiq [Scalable Data Science]}

\author{Ghislain Fourny}
\orcid{0000-0001-8740-8866}
\affiliation{%
  \institution{ETH Zurich}
  \streetaddress{Stampenbachstrasse 114}
  \city{Zurich}
  \country{Switzerland}
  \postcode{8057}
}
\email{ghislain.fourny@inf.ethz.ch}

\author{David Dao}
\affiliation{%
  \institution{ETH Zurich}
  \city{Zurich}
  \country{Switzerland}
  \postcode{8057}
}
\email{david.dao@inf.ethz.ch}

\author{Can Berker Cikis}
\authornote{The work was performed during his studies at ETH Zurich.}
\affiliation{%
  \institution{(unaffiliated)}
  \city{Zurich}
  \country{Switzerland}
  \postcode{8057}
}
\email{canberkerwork@gmail.com}

\author{Ce Zhang}
\affiliation{%
  \institution{ETH Zurich}
  \streetaddress{Stampenbachstrasse 114}
  \city{Zurich}
  \country{Switzerland}
  \postcode{8057}
}
\email{ce.zhang@inf.ethz.ch}

\author{Gustavo Alonso}
\affiliation{%
  \institution{ETH Zurich}
  \streetaddress{Stampenbachstrasse 114}
  \city{Zurich}
  \country{Switzerland}
  \postcode{8057}
}
\email{alonso@inf.ethz.ch}
\email{}
\email{}

\date{January 2021}

\makeatletter
\def\PYG@reset{\let\PYG@it=\relax \let\PYG@bf=\relax%
    \let\PYG@ul=\relax \let\PYG@tc=\relax%
    \let\PYG@bc=\relax \let\PYG@ff=\relax}
\def\PYG@tok#1{\csname PYG@tok@#1\endcsname}
\def\PYG@toks#1+{\ifx\relax#1\empty\else%
    \PYG@tok{#1}\expandafter\PYG@toks\fi}
\def\PYG@do#1{\PYG@bc{\PYG@tc{\PYG@ul{%
    \PYG@it{\PYG@bf{\PYG@ff{#1}}}}}}}
\def\PYG#1#2{\PYG@reset\PYG@toks#1+\relax+\PYG@do{#2}}

\@namedef{PYG@tok@w}{\def\PYG@tc##1{\textcolor[rgb]{0.73,0.73,0.73}{##1}}}
\@namedef{PYG@tok@c}{\let\PYG@it=\textit\def\PYG@tc##1{\textcolor[rgb]{0.25,0.50,0.50}{##1}}}
\@namedef{PYG@tok@cp}{\def\PYG@tc##1{\textcolor[rgb]{0.74,0.48,0.00}{##1}}}
\@namedef{PYG@tok@k}{\let\PYG@bf=\textbf\def\PYG@tc##1{\textcolor[rgb]{0.00,0.50,0.00}{##1}}}
\@namedef{PYG@tok@kp}{\def\PYG@tc##1{\textcolor[rgb]{0.00,0.50,0.00}{##1}}}
\@namedef{PYG@tok@kt}{\def\PYG@tc##1{\textcolor[rgb]{0.69,0.00,0.25}{##1}}}
\@namedef{PYG@tok@o}{\def\PYG@tc##1{\textcolor[rgb]{0.40,0.40,0.40}{##1}}}
\@namedef{PYG@tok@ow}{\let\PYG@bf=\textbf\def\PYG@tc##1{\textcolor[rgb]{0.67,0.13,1.00}{##1}}}
\@namedef{PYG@tok@nb}{\def\PYG@tc##1{\textcolor[rgb]{0.00,0.50,0.00}{##1}}}
\@namedef{PYG@tok@nf}{\def\PYG@tc##1{\textcolor[rgb]{0.00,0.00,1.00}{##1}}}
\@namedef{PYG@tok@nc}{\let\PYG@bf=\textbf\def\PYG@tc##1{\textcolor[rgb]{0.00,0.00,1.00}{##1}}}
\@namedef{PYG@tok@nn}{\let\PYG@bf=\textbf\def\PYG@tc##1{\textcolor[rgb]{0.00,0.00,1.00}{##1}}}
\@namedef{PYG@tok@ne}{\let\PYG@bf=\textbf\def\PYG@tc##1{\textcolor[rgb]{0.82,0.25,0.23}{##1}}}
\@namedef{PYG@tok@nv}{\def\PYG@tc##1{\textcolor[rgb]{0.10,0.09,0.49}{##1}}}
\@namedef{PYG@tok@no}{\def\PYG@tc##1{\textcolor[rgb]{0.53,0.00,0.00}{##1}}}
\@namedef{PYG@tok@nl}{\def\PYG@tc##1{\textcolor[rgb]{0.63,0.63,0.00}{##1}}}
\@namedef{PYG@tok@ni}{\let\PYG@bf=\textbf\def\PYG@tc##1{\textcolor[rgb]{0.60,0.60,0.60}{##1}}}
\@namedef{PYG@tok@na}{\def\PYG@tc##1{\textcolor[rgb]{0.49,0.56,0.16}{##1}}}
\@namedef{PYG@tok@nt}{\let\PYG@bf=\textbf\def\PYG@tc##1{\textcolor[rgb]{0.00,0.50,0.00}{##1}}}
\@namedef{PYG@tok@nd}{\def\PYG@tc##1{\textcolor[rgb]{0.67,0.13,1.00}{##1}}}
\@namedef{PYG@tok@s}{\def\PYG@tc##1{\textcolor[rgb]{0.73,0.13,0.13}{##1}}}
\@namedef{PYG@tok@sd}{\let\PYG@it=\textit\def\PYG@tc##1{\textcolor[rgb]{0.73,0.13,0.13}{##1}}}
\@namedef{PYG@tok@si}{\let\PYG@bf=\textbf\def\PYG@tc##1{\textcolor[rgb]{0.73,0.40,0.53}{##1}}}
\@namedef{PYG@tok@se}{\let\PYG@bf=\textbf\def\PYG@tc##1{\textcolor[rgb]{0.73,0.40,0.13}{##1}}}
\@namedef{PYG@tok@sr}{\def\PYG@tc##1{\textcolor[rgb]{0.73,0.40,0.53}{##1}}}
\@namedef{PYG@tok@ss}{\def\PYG@tc##1{\textcolor[rgb]{0.10,0.09,0.49}{##1}}}
\@namedef{PYG@tok@sx}{\def\PYG@tc##1{\textcolor[rgb]{0.00,0.50,0.00}{##1}}}
\@namedef{PYG@tok@m}{\def\PYG@tc##1{\textcolor[rgb]{0.40,0.40,0.40}{##1}}}
\@namedef{PYG@tok@gh}{\let\PYG@bf=\textbf\def\PYG@tc##1{\textcolor[rgb]{0.00,0.00,0.50}{##1}}}
\@namedef{PYG@tok@gu}{\let\PYG@bf=\textbf\def\PYG@tc##1{\textcolor[rgb]{0.50,0.00,0.50}{##1}}}
\@namedef{PYG@tok@gd}{\def\PYG@tc##1{\textcolor[rgb]{0.63,0.00,0.00}{##1}}}
\@namedef{PYG@tok@gi}{\def\PYG@tc##1{\textcolor[rgb]{0.00,0.63,0.00}{##1}}}
\@namedef{PYG@tok@gr}{\def\PYG@tc##1{\textcolor[rgb]{1.00,0.00,0.00}{##1}}}
\@namedef{PYG@tok@ge}{\let\PYG@it=\textit}
\@namedef{PYG@tok@gs}{\let\PYG@bf=\textbf}
\@namedef{PYG@tok@gp}{\let\PYG@bf=\textbf\def\PYG@tc##1{\textcolor[rgb]{0.00,0.00,0.50}{##1}}}
\@namedef{PYG@tok@go}{\def\PYG@tc##1{\textcolor[rgb]{0.53,0.53,0.53}{##1}}}
\@namedef{PYG@tok@gt}{\def\PYG@tc##1{\textcolor[rgb]{0.00,0.27,0.87}{##1}}}
\@namedef{PYG@tok@err}{\def\PYG@bc##1{{\setlength{\fboxsep}{\string -\fboxrule}\fcolorbox[rgb]{1.00,0.00,0.00}{1,1,1}{\strut ##1}}}}
\@namedef{PYG@tok@kc}{\let\PYG@bf=\textbf\def\PYG@tc##1{\textcolor[rgb]{0.00,0.50,0.00}{##1}}}
\@namedef{PYG@tok@kd}{\let\PYG@bf=\textbf\def\PYG@tc##1{\textcolor[rgb]{0.00,0.50,0.00}{##1}}}
\@namedef{PYG@tok@kn}{\let\PYG@bf=\textbf\def\PYG@tc##1{\textcolor[rgb]{0.00,0.50,0.00}{##1}}}
\@namedef{PYG@tok@kr}{\let\PYG@bf=\textbf\def\PYG@tc##1{\textcolor[rgb]{0.00,0.50,0.00}{##1}}}
\@namedef{PYG@tok@bp}{\def\PYG@tc##1{\textcolor[rgb]{0.00,0.50,0.00}{##1}}}
\@namedef{PYG@tok@fm}{\def\PYG@tc##1{\textcolor[rgb]{0.00,0.00,1.00}{##1}}}
\@namedef{PYG@tok@vc}{\def\PYG@tc##1{\textcolor[rgb]{0.10,0.09,0.49}{##1}}}
\@namedef{PYG@tok@vg}{\def\PYG@tc##1{\textcolor[rgb]{0.10,0.09,0.49}{##1}}}
\@namedef{PYG@tok@vi}{\def\PYG@tc##1{\textcolor[rgb]{0.10,0.09,0.49}{##1}}}
\@namedef{PYG@tok@vm}{\def\PYG@tc##1{\textcolor[rgb]{0.10,0.09,0.49}{##1}}}
\@namedef{PYG@tok@sa}{\def\PYG@tc##1{\textcolor[rgb]{0.73,0.13,0.13}{##1}}}
\@namedef{PYG@tok@sb}{\def\PYG@tc##1{\textcolor[rgb]{0.73,0.13,0.13}{##1}}}
\@namedef{PYG@tok@sc}{\def\PYG@tc##1{\textcolor[rgb]{0.73,0.13,0.13}{##1}}}
\@namedef{PYG@tok@dl}{\def\PYG@tc##1{\textcolor[rgb]{0.73,0.13,0.13}{##1}}}
\@namedef{PYG@tok@s2}{\def\PYG@tc##1{\textcolor[rgb]{0.73,0.13,0.13}{##1}}}
\@namedef{PYG@tok@sh}{\def\PYG@tc##1{\textcolor[rgb]{0.73,0.13,0.13}{##1}}}
\@namedef{PYG@tok@s1}{\def\PYG@tc##1{\textcolor[rgb]{0.73,0.13,0.13}{##1}}}
\@namedef{PYG@tok@mb}{\def\PYG@tc##1{\textcolor[rgb]{0.40,0.40,0.40}{##1}}}
\@namedef{PYG@tok@mf}{\def\PYG@tc##1{\textcolor[rgb]{0.40,0.40,0.40}{##1}}}
\@namedef{PYG@tok@mh}{\def\PYG@tc##1{\textcolor[rgb]{0.40,0.40,0.40}{##1}}}
\@namedef{PYG@tok@mi}{\def\PYG@tc##1{\textcolor[rgb]{0.40,0.40,0.40}{##1}}}
\@namedef{PYG@tok@il}{\def\PYG@tc##1{\textcolor[rgb]{0.40,0.40,0.40}{##1}}}
\@namedef{PYG@tok@mo}{\def\PYG@tc##1{\textcolor[rgb]{0.40,0.40,0.40}{##1}}}
\@namedef{PYG@tok@ch}{\let\PYG@it=\textit\def\PYG@tc##1{\textcolor[rgb]{0.25,0.50,0.50}{##1}}}
\@namedef{PYG@tok@cm}{\let\PYG@it=\textit\def\PYG@tc##1{\textcolor[rgb]{0.25,0.50,0.50}{##1}}}
\@namedef{PYG@tok@cpf}{\let\PYG@it=\textit\def\PYG@tc##1{\textcolor[rgb]{0.25,0.50,0.50}{##1}}}
\@namedef{PYG@tok@c1}{\let\PYG@it=\textit\def\PYG@tc##1{\textcolor[rgb]{0.25,0.50,0.50}{##1}}}
\@namedef{PYG@tok@cs}{\let\PYG@it=\textit\def\PYG@tc##1{\textcolor[rgb]{0.25,0.50,0.50}{##1}}}

\def\PYGZbs{\char`\\}
\def\PYGZus{\char`\_}
\def\PYGZob{\char`\{}
\def\PYGZcb{\char`\}}
\def\PYGZca{\char`\^}
\def\PYGZam{\char`\&}
\def\PYGZlt{\char`\<}
\def\PYGZgt{\char`\>}
\def\PYGZsh{\char`\#}
\def\PYGZpc{\char`\%}
\def\PYGZdl{\char`\$}
\def\PYGZhy{\char`\-}
\def\PYGZsq{\char`\'}
\def\PYGZdq{\char`\"}
\def\PYGZti{\char`\~}
\def\PYGZat{@}
\def\PYGZlb{[}
\def\PYGZrb{]}
\makeatother

\makeatletter
\def\PYGdefault@reset{\let\PYGdefault@it=\relax \let\PYGdefault@bf=\relax%
    \let\PYGdefault@ul=\relax \let\PYGdefault@tc=\relax%
    \let\PYGdefault@bc=\relax \let\PYGdefault@ff=\relax}
\def\PYGdefault@tok#1{\csname PYGdefault@tok@#1\endcsname}
\def\PYGdefault@toks#1+{\ifx\relax#1\empty\else%
    \PYGdefault@tok{#1}\expandafter\PYGdefault@toks\fi}
\def\PYGdefault@do#1{\PYGdefault@bc{\PYGdefault@tc{\PYGdefault@ul{%
    \PYGdefault@it{\PYGdefault@bf{\PYGdefault@ff{#1}}}}}}}
\def\PYGdefault#1#2{\PYGdefault@reset\PYGdefault@toks#1+\relax+\PYGdefault@do{#2}}

\@namedef{PYGdefault@tok@w}{\def\PYGdefault@tc##1{\textcolor[rgb]{0.73,0.73,0.73}{##1}}}
\@namedef{PYGdefault@tok@c}{\let\PYGdefault@it=\textit\def\PYGdefault@tc##1{\textcolor[rgb]{0.25,0.50,0.50}{##1}}}
\@namedef{PYGdefault@tok@cp}{\def\PYGdefault@tc##1{\textcolor[rgb]{0.74,0.48,0.00}{##1}}}
\@namedef{PYGdefault@tok@k}{\let\PYGdefault@bf=\textbf\def\PYGdefault@tc##1{\textcolor[rgb]{0.00,0.50,0.00}{##1}}}
\@namedef{PYGdefault@tok@kp}{\def\PYGdefault@tc##1{\textcolor[rgb]{0.00,0.50,0.00}{##1}}}
\@namedef{PYGdefault@tok@kt}{\def\PYGdefault@tc##1{\textcolor[rgb]{0.69,0.00,0.25}{##1}}}
\@namedef{PYGdefault@tok@o}{\def\PYGdefault@tc##1{\textcolor[rgb]{0.40,0.40,0.40}{##1}}}
\@namedef{PYGdefault@tok@ow}{\let\PYGdefault@bf=\textbf\def\PYGdefault@tc##1{\textcolor[rgb]{0.67,0.13,1.00}{##1}}}
\@namedef{PYGdefault@tok@nb}{\def\PYGdefault@tc##1{\textcolor[rgb]{0.00,0.50,0.00}{##1}}}
\@namedef{PYGdefault@tok@nf}{\def\PYGdefault@tc##1{\textcolor[rgb]{0.00,0.00,1.00}{##1}}}
\@namedef{PYGdefault@tok@nc}{\let\PYGdefault@bf=\textbf\def\PYGdefault@tc##1{\textcolor[rgb]{0.00,0.00,1.00}{##1}}}
\@namedef{PYGdefault@tok@nn}{\let\PYGdefault@bf=\textbf\def\PYGdefault@tc##1{\textcolor[rgb]{0.00,0.00,1.00}{##1}}}
\@namedef{PYGdefault@tok@ne}{\let\PYGdefault@bf=\textbf\def\PYGdefault@tc##1{\textcolor[rgb]{0.82,0.25,0.23}{##1}}}
\@namedef{PYGdefault@tok@nv}{\def\PYGdefault@tc##1{\textcolor[rgb]{0.10,0.09,0.49}{##1}}}
\@namedef{PYGdefault@tok@no}{\def\PYGdefault@tc##1{\textcolor[rgb]{0.53,0.00,0.00}{##1}}}
\@namedef{PYGdefault@tok@nl}{\def\PYGdefault@tc##1{\textcolor[rgb]{0.63,0.63,0.00}{##1}}}
\@namedef{PYGdefault@tok@ni}{\let\PYGdefault@bf=\textbf\def\PYGdefault@tc##1{\textcolor[rgb]{0.60,0.60,0.60}{##1}}}
\@namedef{PYGdefault@tok@na}{\def\PYGdefault@tc##1{\textcolor[rgb]{0.49,0.56,0.16}{##1}}}
\@namedef{PYGdefault@tok@nt}{\let\PYGdefault@bf=\textbf\def\PYGdefault@tc##1{\textcolor[rgb]{0.00,0.50,0.00}{##1}}}
\@namedef{PYGdefault@tok@nd}{\def\PYGdefault@tc##1{\textcolor[rgb]{0.67,0.13,1.00}{##1}}}
\@namedef{PYGdefault@tok@s}{\def\PYGdefault@tc##1{\textcolor[rgb]{0.73,0.13,0.13}{##1}}}
\@namedef{PYGdefault@tok@sd}{\let\PYGdefault@it=\textit\def\PYGdefault@tc##1{\textcolor[rgb]{0.73,0.13,0.13}{##1}}}
\@namedef{PYGdefault@tok@si}{\let\PYGdefault@bf=\textbf\def\PYGdefault@tc##1{\textcolor[rgb]{0.73,0.40,0.53}{##1}}}
\@namedef{PYGdefault@tok@se}{\let\PYGdefault@bf=\textbf\def\PYGdefault@tc##1{\textcolor[rgb]{0.73,0.40,0.13}{##1}}}
\@namedef{PYGdefault@tok@sr}{\def\PYGdefault@tc##1{\textcolor[rgb]{0.73,0.40,0.53}{##1}}}
\@namedef{PYGdefault@tok@ss}{\def\PYGdefault@tc##1{\textcolor[rgb]{0.10,0.09,0.49}{##1}}}
\@namedef{PYGdefault@tok@sx}{\def\PYGdefault@tc##1{\textcolor[rgb]{0.00,0.50,0.00}{##1}}}
\@namedef{PYGdefault@tok@m}{\def\PYGdefault@tc##1{\textcolor[rgb]{0.40,0.40,0.40}{##1}}}
\@namedef{PYGdefault@tok@gh}{\let\PYGdefault@bf=\textbf\def\PYGdefault@tc##1{\textcolor[rgb]{0.00,0.00,0.50}{##1}}}
\@namedef{PYGdefault@tok@gu}{\let\PYGdefault@bf=\textbf\def\PYGdefault@tc##1{\textcolor[rgb]{0.50,0.00,0.50}{##1}}}
\@namedef{PYGdefault@tok@gd}{\def\PYGdefault@tc##1{\textcolor[rgb]{0.63,0.00,0.00}{##1}}}
\@namedef{PYGdefault@tok@gi}{\def\PYGdefault@tc##1{\textcolor[rgb]{0.00,0.63,0.00}{##1}}}
\@namedef{PYGdefault@tok@gr}{\def\PYGdefault@tc##1{\textcolor[rgb]{1.00,0.00,0.00}{##1}}}
\@namedef{PYGdefault@tok@ge}{\let\PYGdefault@it=\textit}
\@namedef{PYGdefault@tok@gs}{\let\PYGdefault@bf=\textbf}
\@namedef{PYGdefault@tok@gp}{\let\PYGdefault@bf=\textbf\def\PYGdefault@tc##1{\textcolor[rgb]{0.00,0.00,0.50}{##1}}}
\@namedef{PYGdefault@tok@go}{\def\PYGdefault@tc##1{\textcolor[rgb]{0.53,0.53,0.53}{##1}}}
\@namedef{PYGdefault@tok@gt}{\def\PYGdefault@tc##1{\textcolor[rgb]{0.00,0.27,0.87}{##1}}}
\@namedef{PYGdefault@tok@err}{\def\PYGdefault@bc##1{{\setlength{\fboxsep}{\string -\fboxrule}\fcolorbox[rgb]{1.00,0.00,0.00}{1,1,1}{\strut ##1}}}}
\@namedef{PYGdefault@tok@kc}{\let\PYGdefault@bf=\textbf\def\PYGdefault@tc##1{\textcolor[rgb]{0.00,0.50,0.00}{##1}}}
\@namedef{PYGdefault@tok@kd}{\let\PYGdefault@bf=\textbf\def\PYGdefault@tc##1{\textcolor[rgb]{0.00,0.50,0.00}{##1}}}
\@namedef{PYGdefault@tok@kn}{\let\PYGdefault@bf=\textbf\def\PYGdefault@tc##1{\textcolor[rgb]{0.00,0.50,0.00}{##1}}}
\@namedef{PYGdefault@tok@kr}{\let\PYGdefault@bf=\textbf\def\PYGdefault@tc##1{\textcolor[rgb]{0.00,0.50,0.00}{##1}}}
\@namedef{PYGdefault@tok@bp}{\def\PYGdefault@tc##1{\textcolor[rgb]{0.00,0.50,0.00}{##1}}}
\@namedef{PYGdefault@tok@fm}{\def\PYGdefault@tc##1{\textcolor[rgb]{0.00,0.00,1.00}{##1}}}
\@namedef{PYGdefault@tok@vc}{\def\PYGdefault@tc##1{\textcolor[rgb]{0.10,0.09,0.49}{##1}}}
\@namedef{PYGdefault@tok@vg}{\def\PYGdefault@tc##1{\textcolor[rgb]{0.10,0.09,0.49}{##1}}}
\@namedef{PYGdefault@tok@vi}{\def\PYGdefault@tc##1{\textcolor[rgb]{0.10,0.09,0.49}{##1}}}
\@namedef{PYGdefault@tok@vm}{\def\PYGdefault@tc##1{\textcolor[rgb]{0.10,0.09,0.49}{##1}}}
\@namedef{PYGdefault@tok@sa}{\def\PYGdefault@tc##1{\textcolor[rgb]{0.73,0.13,0.13}{##1}}}
\@namedef{PYGdefault@tok@sb}{\def\PYGdefault@tc##1{\textcolor[rgb]{0.73,0.13,0.13}{##1}}}
\@namedef{PYGdefault@tok@sc}{\def\PYGdefault@tc##1{\textcolor[rgb]{0.73,0.13,0.13}{##1}}}
\@namedef{PYGdefault@tok@dl}{\def\PYGdefault@tc##1{\textcolor[rgb]{0.73,0.13,0.13}{##1}}}
\@namedef{PYGdefault@tok@s2}{\def\PYGdefault@tc##1{\textcolor[rgb]{0.73,0.13,0.13}{##1}}}
\@namedef{PYGdefault@tok@sh}{\def\PYGdefault@tc##1{\textcolor[rgb]{0.73,0.13,0.13}{##1}}}
\@namedef{PYGdefault@tok@s1}{\def\PYGdefault@tc##1{\textcolor[rgb]{0.73,0.13,0.13}{##1}}}
\@namedef{PYGdefault@tok@mb}{\def\PYGdefault@tc##1{\textcolor[rgb]{0.40,0.40,0.40}{##1}}}
\@namedef{PYGdefault@tok@mf}{\def\PYGdefault@tc##1{\textcolor[rgb]{0.40,0.40,0.40}{##1}}}
\@namedef{PYGdefault@tok@mh}{\def\PYGdefault@tc##1{\textcolor[rgb]{0.40,0.40,0.40}{##1}}}
\@namedef{PYGdefault@tok@mi}{\def\PYGdefault@tc##1{\textcolor[rgb]{0.40,0.40,0.40}{##1}}}
\@namedef{PYGdefault@tok@il}{\def\PYGdefault@tc##1{\textcolor[rgb]{0.40,0.40,0.40}{##1}}}
\@namedef{PYGdefault@tok@mo}{\def\PYGdefault@tc##1{\textcolor[rgb]{0.40,0.40,0.40}{##1}}}
\@namedef{PYGdefault@tok@ch}{\let\PYGdefault@it=\textit\def\PYGdefault@tc##1{\textcolor[rgb]{0.25,0.50,0.50}{##1}}}
\@namedef{PYGdefault@tok@cm}{\let\PYGdefault@it=\textit\def\PYGdefault@tc##1{\textcolor[rgb]{0.25,0.50,0.50}{##1}}}
\@namedef{PYGdefault@tok@cpf}{\let\PYGdefault@it=\textit\def\PYGdefault@tc##1{\textcolor[rgb]{0.25,0.50,0.50}{##1}}}
\@namedef{PYGdefault@tok@c1}{\let\PYGdefault@it=\textit\def\PYGdefault@tc##1{\textcolor[rgb]{0.25,0.50,0.50}{##1}}}
\@namedef{PYGdefault@tok@cs}{\let\PYGdefault@it=\textit\def\PYGdefault@tc##1{\textcolor[rgb]{0.25,0.50,0.50}{##1}}}

\def\PYGdefaultZbs{\char`\\}
\def\PYGdefaultZus{\char`\_}
\def\PYGdefaultZob{\char`\{}
\def\PYGdefaultZcb{\char`\}}
\def\PYGdefaultZca{\char`\^}
\def\PYGdefaultZam{\char`\&}
\def\PYGdefaultZlt{\char`\<}
\def\PYGdefaultZgt{\char`\>}
\def\PYGdefaultZsh{\char`\#}
\def\PYGdefaultZpc{\char`\%}
\def\PYGdefaultZdl{\char`\$}
\def\PYGdefaultZhy{\char`\-}
\def\PYGdefaultZsq{\char`\'}
\def\PYGdefaultZdq{\char`\"}
\def\PYGdefaultZti{\char`\~}
\def\PYGdefaultZat{@}
\def\PYGdefaultZlb{[}
\def\PYGdefaultZrb{]}
\makeatother

\begin{abstract}
Lakehouse systems have reached in the past few years unprecedented size and heterogeneity and have been embraced by many industry players. However, they are often difficult to use as they lack the declarative language and optimization possibilities of relational engines. This paper introduces RumbleML, a high-level, declarative library integrated into the RumbleDB engine and with the JSONiq language. RumbleML allows using a single platform for data cleaning, data preparation, training, and inference, as well as management of models and results. It does it using a purely declarative language (JSONiq) for all these tasks and without any performance loss over existing platforms (e.g. Spark).  
The key insights of the design of RumbleML are that training sets, evaluation sets, and test sets can be represented as homogeneous sequences of flat objects; that models can be seamlessly embodied in function items mapping input test sets into prediction-augmented result sets; and that estimators can be seamlessly embodied in function items mapping input training sets to models. We argue that this makes JSONiq a viable and seamless programming language for data lakehouses across all their features, whether database-related or machine-learning-related. While lakehouses bring Machine Learning and Data Wrangling on the same platform, RumbleML also brings them to the same language, JSONiq. In the paper, we present the first prototype and compare its performance to Spark showing the benefit of a huge functionality and productivity gain for cleaning up, normalizing, validating data, feeding it into Machine Learning pipelines, and analyzing the output, all within the same system and language and at scale.
\end{abstract}

\maketitle

\pagestyle{\vldbpagestyle}
\begingroup\small\noindent\raggedright\textbf{PVLDB Reference Format:}\\
\vldbauthors. \vldbtitle. PVLDB, \vldbvolume(\vldbissue): \vldbpages, \vldbyear.\\
\href{https://doi.org/\vldbdoi}{doi:\vldbdoi}
\endgroup
\begingroup
\renewcommand\thefootnote{}\footnote{\noindent
This work is licensed under the Creative Commons BY-NC-ND 4.0 International License. Visit \url{https://creativecommons.org/licenses/by-nc-nd/4.0/} to view a copy of this license. For any use beyond those covered by this license, obtain permission by emailing \href{mailto:info@vldb.org}{info@vldb.org}. Copyright is held by the owner/author(s). Publication rights licensed to the VLDB Endowment. \\
\raggedright Proceedings of the VLDB Endowment, Vol. \vldbvolume, No. \vldbissue\ %
ISSN 2150-8097. \\
\href{https://doi.org/\vldbdoi}{doi:\vldbdoi} \\
}\addtocounter{footnote}{-1}\endgroup

\ifdefempty{\vldbavailabilityurl}{}{
\vspace{.3cm}
\begingroup\small\noindent\raggedright\textbf{PVLDB Artifact Availability:}\\
The source code, data, and/or other artifacts have been made available at \url{\vldbavailabilityurl}.
\endgroup
}

\section{Introduction}
Machine learning has been enjoying ever-increasing popularity in the past years, thanks to the growing availability of data, computing power, and numerous theoretical breakthroughs, such as Generative Adversarial Networks (GANs)~\cite{goodfellow2014}, deep learning~\cite{hinton2007}, and so on. Machine Learning (ML) is slowly becoming an indispensable part of database management systems, whether they are databases, data stores, data warehouses, or data lakes. A new paradigm called \emph{lakehouse} has recently been suggested~\cite{armbrust2021lakehouse} integrating structured and unstructured data management, as well as ML under the same platform. Unfortunately, lakehouses still lack a common language, forcing users to resort to different languages and systems for different tasks: data cleaning and preparation, training and inference, as well as model and result management.

ML workflows frequently follow a conventional pattern. Firstly, all available data is split into two overarching groups: the training data and the test data. The training data is used to train a model, that is, identify the important features in data and assign weights to them. These weights serve as the basis for predictions. The trained model's performance is evaluated by applying it to the test data. The goal here is to optimize the model to generate correct predictions on any given dataset while avoiding over-fitting on the training data. Obtaining a model with satisfactory performance may take a couple of iterations. Then, it is ready to be applied to production data to generate predictions.

ML still faces challenges that impede an even broader use \cite{ilyas2021}\cite{Agrawal2019}. First, traditional ML expects clean and highly-structured data input. Unfortunately, data is often heterogeneous, nested, and messy in the real world. It requires preparation, clean up and curation, as well as normalization. These are usually not well addressed in the design of ML software, and ML practitioners frequently report data preparation to be their biggest challenge and most time consuming~\cite{press2016}.

\begin{figure}

\begin{minted}[breaklines]{xquery}
declare function local:convert($input)
{
 annotate(
  for $l in unparsed-text-lines($input)
  let $tokens := tokenize ($l, " ")
  let $left := head($tokens)
  let $right := tail($tokens)
  let $label := if (contains($left, "indoor"))
         then 0
         else 1
  let $features := {|
	for $i at $p in $right
	return { string($p) : $i }
  |}
  return { "label" : $label, "features" : $features },
  {
    "label" : "string",
    "features" : {|
      for $i in 1 to 4096
      return { string($i) : "double" }
    |}
  }
 )
};
let $training-data := local:convert($training-input)
let $test-data := local:convert($test-input)
let $vector-assembler := get-transformer(
  "VectorAssembler",
  {
    "inputCols" : ["features"],
    "outputCol" : "transformedFeatures"
  }
)
let $linearsvc := get-estimator(
  "LinearSVC",
  {
    "featuresCol": "transformedFeatures",
    "maxIter": 5
  }
)
let $pipeline := get-estimator(
  "Pipeline",
  {
    "stages": [$vector-assembler, $linearsvc]
  }
)
let $pip := $pipeline($training-data, {})
let $prediction := $pip($test-data, {})
let $total := count($prediction)
return count($prediction[$$.label eq $$.prediction])
        div $total
\end{minted}

\caption{An entire pipeline, from cleaning the messy text input to computing accuracy, entirely written in JSONiq.}
\label{fig-JSONiq-cleanup}
\end{figure}

Second, most ML systems require the user to code at a low level with an awareness of the physical data layout. These frameworks are commonly built-in imperative languages and do not offer much help with data cleaning tasks. This situation leaves the user with two options: either working in the same programming environment and implementing data cleaning scripts in these iterative languages or switching to a different programming environment. Users generally opt for the former option, e.g., with Python. The scripts iterate over the data and perform conditional formatting that resembles a chain of if/else statements. This practice has a major drawback in terms of user productivity: Implementing complicated value mappings in imperative scripts increases development and maintenance costs as they may require advanced programming knowledge and can be error-prone.

Third, many systems such as TensorFlow~\cite{abadi2016tensorflow},  Weka~\cite{hall2009weka}, PyTorch~\cite{paszke2017automatic} require advanced knowledge of ML and are only accessible to ML experts. There is awareness in the ML community for the need for a declarative ML approach~\cite{boehm2016systemml}, both in the form of declarative ML tasks and in the form of declarative ML algorithms.

Fourth, most tools run locally and are thus restricted in the amount of data that they can process. Among the tools that scale, spark.ml is often cited as the one enjoying the most popularity in the open-source world.

As it turns out, these challenges are very well addressed by the database community when it comes solely to data management \cite{Abouzied2018} thanks to the widely adopted SQL language~\cite{date1987guide} for tabular data, and a newer generation of languages such as JSONiq for nested, heterogeneous datasets~\cite{muller2020rumble}. In this paper, we suggest taking this avenue further by taking a programming language, JSONiq, to a full lakehouse system. The training, evaluation, and use of ML models are seamlessly integrated into the database ecosystem rather than the other way round. We argue that JSONiq is well suited for a lakehouse, as it is capable at the same time of (i) performing well on highly structured datasets, (ii) preparing, cleaning up, curating, and normalizing the data, and (iii) supporting ML training and evaluation via its higher-order function feature as shown in this paper. An example JSONiq program that does all of this is shown in Figure \ref{fig-JSONiq-cleanup}.

To prove that such is the case, we built a first prototype integrating the functionality offered by the spark.ml library into the distributed RumbleDB engine. RumbleDB is designed to improve user productivity in parallel processing through the utilization of the declarative JSONiq language rather than the iterative APIs of Spark. RumbleML extends this notion into the ML domain by exposing the capabilities of spark.ml. By delegating the computation with as little overhead as possible, RumbleML has no loss of performance in comparison to spark.ml. The main design goal of RumbleML is not to be the fastest ML framework but the most productive for the user. To implement RumbleML, we did not design a new ML framework from scratch, but rather, we exposed spark.ml as faithfully as possible. We also did not design a new query language or data processing engine, but instead, we extended the existing RumbleDB engine with its JSONiq language.

The successful implementation of RumbleML demonstrates the feasibility of seamless integration of the spectrum of lakehouse features under the same declarative, functional, and high-level query language, hiding the underlying gluing from the user. While in this paper we focus on declarative ML tasks, RumbleML is also forward compatible with declarative ML algorithms, as function items can be implemented by users as well and need not be restricted to spark.ml functionality. Our approach opens the avenue for research on optimizing the execution further and automatically tuning ML parameters within this data-independent framework.

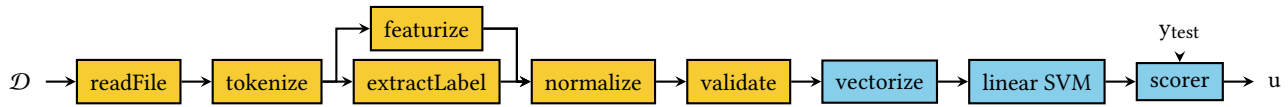
\begin{figure*}[!ht]
\centering
\begin{tikzpicture}[align=center, node distance=1mm and 4mm, line width=0.8pt] 
    \tikzstyle{free} = [inner sep=5pt]
    \tikzstyle{pbox} = [draw, rectangle, fill=myyellow, inner sep=5pt]
    \tikzstyle{mbox} = [draw, rectangle, fill=myblue, inner sep=5pt]
    
    \node[free] (dataset) {$\mathcal{D}$};
    \node[pbox] (read) [right=of dataset] {$\mathrm{readFile}$};
    \node[pbox] (tokenizer) [right=of read] {$\mathrm{tokenize}$};
    \node[pbox] (extractLabel) [right=of tokenizer] {$\mathrm{extractLabel}$};
    \node[pbox] (featurize) [above=of extractLabel] {$\mathrm{featurize}$};
    \node[pbox] (normalize) [right=of extractLabel] {$\mathrm{normalize}$};
    \node[pbox] (validate) [right=of normalize] {$\mathrm{validate}$};
    \node[mbox] (vectorize) [right=of validate] {$\mathrm{vectorize}$};
    \node[mbox] (linearSVC) [right=of vectorize] {$\mathrm{linear\ SVM}$};
    \node[mbox] (scorer) [right=of linearSVC] {$\mathrm{scorer}$};
    \node[free] (y_test) [above=of scorer]
    {$\mathrm{y_{test}}$};
    \node[free] (score) [right=of scorer]
    {$\mathrm{u}$};
    
    \draw[-stealth] (dataset) -- (read);
    \draw[-stealth] (read) -- (tokenizer);
    \draw[-stealth] (tokenizer) -| ($(extractLabel)-(1.2,0)$) |- (extractLabel);
    \draw[-stealth] (tokenizer) -| ($(extractLabel)-(1.2,0)$) |- (featurize);
    \draw[-stealth] (featurize) -| ($(extractLabel)+(1.2,0)$) |- (normalize);
    \draw[-stealth] (featurize) -| ($(extractLabel)+(1.2,0)$) |- (normalize);
    \draw[-stealth] (extractLabel) -- (normalize);
    \draw[-stealth] (normalize) -- (validate);
    \draw[-stealth] (validate) -- (vectorize);
    \draw[-stealth] (vectorize) -- (linearSVC);
    \draw[-stealth] (linearSVC) -- (scorer);
    \draw[-stealth] (y_test) -- (scorer);
    \draw[-stealth] (scorer) -- (score);
\end{tikzpicture}

\caption{A schematic depicting the RumbleML end-to-end pipeline corresponding to the example code in Figure \ref{fig-JSONiq-cleanup}. The pipeline starts with raw, unstructured data (see Figure \ref{fig-messy-dataset}) that it cleans, normalizes and validates into a DataFrame (yellow). After the data cleaning stage, a linear support vector model is trained and evaluated (blue)} 
\label{fig:summary_fig1}
\end{figure*}

\section{Background}
\noindent\textbf{Spark.}~~%
Apache Spark~\cite{spark} is an open-source, parallel processing framework that competently scales for big data applications. Spark optimizes query execution on top of its fault-tolerant and distributed nature with, e.g., in-memory caching, reduced I/O requirements. On the logical level, Spark primarily manipulates Resilient Distributed Datasets (RDD), which are collections of arbitrary values. Unlike MapReduce, the data flow takes the generic form of Directed Acyclic Graphs (DAG). On the physical level, the DAG is evaluated with parallelism and batch processing on large clusters. Spark covers a wide range of workloads such as batch applications, iterative algorithms, interactive queries, streaming, and graph processing.

\noindent\textbf{Spark SQL.}~~%
Spark also offers a higher abstraction and better performance for highly structured data via DataFrames and the language Spark SQL~\cite{10.1145/2934664}. Unlike an RDD, a DataFrame is a collection of rows that all share the same schema, allowing for recursively nested arrays and structs, similar to Pandas~\cite{mckinney2011pandas}. Physically, the data is stored in memory in a columnar format. The schema is exploited with the Catalyst engine for cost-based optimizations and code generation for query execution. 

\noindent\textbf{spark.ml.}~~%
Spark's generic DAG-based model also supports iterative computations and is compatible with ML algorithms as they are typically iterative. Spark supports ML with the spark.ml library~~\cite{sparkml}. In addition to many ML algorithms, it also offers statistics, optimization, and linear-algebra-related primitives. spark.ml aims to make ML practical, scalable, and easy to use. At a high level, spark.ml comes out of the box with (i) tools for constructing, evaluating, and tuning ML pipelines, (ii) Common ML algorithms such as classification, regression, clustering, and collaborative filtering, (iii) feature extraction, transformation, dimensionality reduction, and selection, (iv) persistence for algorithms, models, and pipelines, (v) utilities such as linear algebra, statistics, data handling~\cite{sparkml}.

\begin{figure}[b]
\centering
\includegraphics[width=\columnwidth]{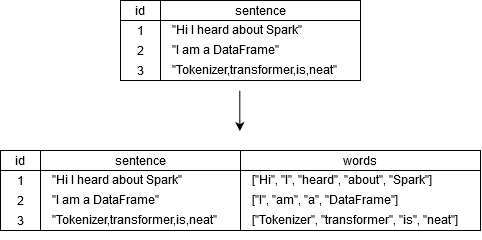}
\caption{A tokenizer is a transformer that adds a column containing arrays of tokens obtained from another column.}
\label{fig-tokenizer}
\end{figure}

spark.ml offers a high-level, uniform API for ML algorithms and featurization. This facilitates the architecture of complete workflows while making it easier and simpler to comprehend how spark.ml components fit together. The main components of spark.ml's pipeline API are (i) DataFrames: spark.ml manipulates ML datasets as structured DataFrames, and a DataFrame can have different columns storing text, feature vectors, true labels, and predictions. spark.ml also extends the DataFrame type system with sparse and dense vectors; (ii)
transformers, which are off-the-shelf algorithms that transform a DataFrame into another DataFrame. E.g., an ML model is a Transformer that extends a DataFrame with predictions. (iii) estimators, which are algorithms that are fit on a DataFrame to produce a fitted model exposed as a transformer; (iv) pipelines, which chain multiple transformers and estimators into an ML workflow; and (v) parameters that tune estimators and transformers with a unified API. An example with the Tokenizer transformers is shown in Figure \ref{fig-tokenizer}.

\noindent\textbf{JSONiq.}
JSONiq is a declarative, functional query language that manipulates nested and/or heterogeneous datasets seamlessly. The JSONiq data model (JDM) is based on every value returned by an expression being a sequence of items. Items can be atomic values, objects, arrays, or function items. Sequences can be heterogeneous or homogeneous, for example, objects valid against the same schema. FLWOR expressions support the entire relational algebra, generalizing SQL's SELECT FROM WHERE to denormalized data. JSONiq is taught at several universities \cite{grust2021}\cite{sunderraman2021}\cite{jsoniq} and is an ideal candidate to easily clean up, structure, normalize and validate messy data into DataFrames~.

\noindent\textbf{RumbleDB.}
RumbleDB~\cite{muller2020rumble} is a querying engine that interfaces the power of Apache Spark with the convenience and versatility of the JSONiq language. Spark SQL only supports structured data, leaving users to use lower-level RDDs for less structured datasets. RumbleDB fixes this with a high-level approach across the entire data normalization spectrum, and both at small and large scales.

\section{Data preparation with JSONiq}
\label{section-data-cleaning}
The biggest challenge met by ML practitioners is data preparation \cite{press2016}\cite{sadiq2017} \cite{chu2016data}. While there exist libraries for data cleaning and preparation \cite{Krishnan2015} \cite{olson2021} \cite{Caveness2020} \cite{rekatsinas2017holoclean} \cite{rezig2021horizon} \cite{krishnan2016activeclean} \cite{chu2016qualitative} \cite{muller2021papers} \cite{geerts2013llunatic} \cite{prokoshyna2015combining}, they often require the data to already be available in tabular form and perform tasks such as filling missing values, identity matching, and so on. However, in reality, messy datasets need not be in tabular form and will not be readily available in DataFrames (e.g., the Git Archive dataset \cite{markovtsev2018}). With RumbleML, the data preparation pipeline is supported as early as with nested, heterogeneous datasets, and even starting with fully unstructured (text) datasets as can be seen in the pipeline shown in Figure \ref{fig:summary_fig1}.

In this section, we show that JSONiq is capable of processing messy datasets (in the sense of nested, heterogeneous, and/or textual) such as JSON Lines files \cite{long2021}, normalizing them, and validating them into structured DataFrames, feeding these DataFrames as input training sets and test sets to ML pipelines, and computing metrics to evaluate the output. In other systems, this requires different tools \cite{microsoft2021}\cite{dong2018data}, because trying to do all of this in the same program is very cumbersome due to the many impedance mismatches. A benchmark \cite{graur2022} was also performed to show the limitations and shortcomings of other query languages and APIs, with a focus on nestedness and a use case in high-energy physics.

We show a messy sample dataset in Figure \ref{fig-messy-dataset}. This dataset is only implicitly structured and requires processing and conversion to a more appropriate, DataFrame-based format. In the current state, this is typically done with a complex Python script. An example of the way it could look like when converted to LibSVM is shown in Figure \ref{fig-cleaned-dataset}. The libSVM file can then be opened by Spark as a DataFrame and fed into spark.ml pipelines.

With RumbleDB this all can be achieved with a single JSONiq program, the one shown in Figure \ref{fig-JSONiq-cleanup}.

\begin{figure}
\begin{minted}[breaklines]{text}
animal:0.7420,outdoor:0.9710,pet:0.6130,white:0.6790 -4.893 -3.803 -25.799 -34.55 -6.622 -13.547 ...
animal:0.1234,indoor:0.3413,pet:0.6130,black:0.87534 -8.311 15.133 2.973 -25.972 -11.422 -0.067 ...
\end{minted}
\caption{Example of unprocessed text data used to generate the $Clustered$ $YFCC$ dataset.}
\label{fig-messy-dataset}
\end{figure}

\begin{figure}
\begin{minted}[breaklines]{text}
0 1:-4.893 2:-3.803 3:-25.799 4:-34.55 5:-6.622 6:-13.547 ...
1 1:-8.311 2:15.133 3:2.973 4:-25.972 5:-11.422 6:-0.067 ...
\end{minted}
\caption{Example of the resulting LibSVM format for YFCC.}
\label{fig-cleaned-dataset}
\end{figure}

DataFrames have become very popular in Data Science APIs, not only in Spark but also in Python (e.g., Pandas). For example, ``pandas'' ranks 32nd among StackOverflow tags at the time of writing with 226,000+ questions \cite{stackoverflowtags}, and ``dataframe'' 66th with 109,000+ questions. They can be seen as an extension of the relational model that supports nested data. DataFrames, like relational tables, have named columns. Unlike relational tables, however, a DataFrame column type need not be atomic (first normal form), but can recursively be an array type whose elements are of any DataFrame type; or a struct type whose keys are strings and whose values are also arbitrary DataFrame types. While DataFrames support denormalized data because of nestedness, they do not support heterogeneity, i.e., values of different types in the same column or array. As a consequence, many datasets, even if they are semi-structured, are too messy to be opened as a DataFrame. For example, datasets in the JSON format where the fields might not always have the same type, be missing, or extraneous, can result in many string columns containing serialized objects and arrays, pushing the burden onto the user to parse it back selectively at a low level.

Note that DataFrame support in other implementations than Apache Spark might behave differently, for example, in Snowflake \cite{Dageville2016}, DataFrames support heterogeneity with the VARIANT type to some extent, but there is less support for nestedness (only for JSON types). DataFrames can thus be seen as a special case of sequences of objects, where this sequence of objects must have some structure, e.g., must have been validated against a DataFrame schema.

\begin{figure}[H]
\centering
\includegraphics[width=\columnwidth]{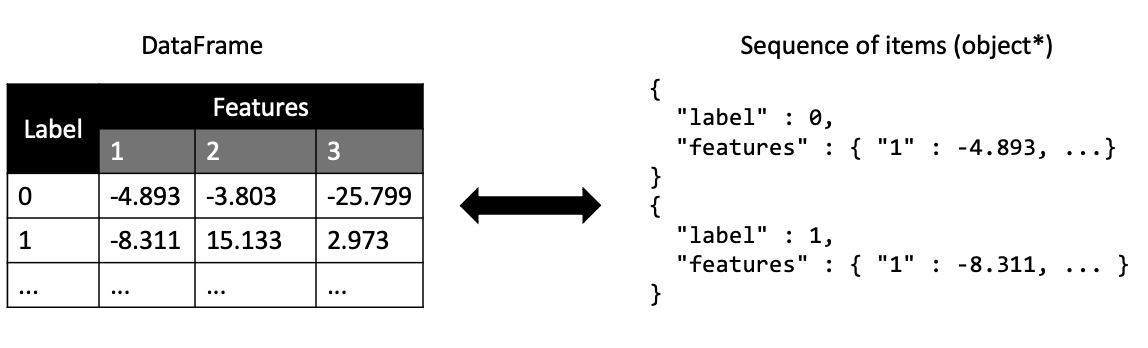}
\caption{Mapping a DataFrame to a sequence of object*. The data corresponds to that processed in Figure \ref{fig-JSONiq-cleanup}, and was cleaned up from the messy textual data shown in Figure \ref{fig-messy-dataset}.}
\label{fig:RumbleML_DFToObjectStar}
\end{figure}
An example showing the conversion between these two analogous representations is given in Figure \ref{fig:RumbleML_DFToObjectStar}. The sequence ({"foo":1}, {"foo" : [ "bar" ]}) would, however, not have any DataFrame equivalent because the field foo is associated with an integer in the first object, and with an array in the second object.

RumbleDB leverages this mapping in a way seamless and transparent to the user, as follows. JSONiq's sequences of items at the logical level have five different implementations at the physical level: (i) directly returning an Item via a Java method call, (ii) volcano-style iteration on Items, (iii) Spark RDD of Items, (iv) DataFrames and (v) conversion to native Spark SQL queries.
DataFrames efficiently implement the special case where the sequence of items is a homogeneous sequence of objects that are all valid against the same schema. ML tasks, however, only work on DataFrames, i.e., on a homogeneous sequence of valid objects. Thus, in JSONiq, cleaning up and preparing the data corresponds to using the language to transform the original messy sequence of items to a fully structured sequence of valid objects. RumbleDB will automatically understand that it can implement the latter internally as a DataFrame.

Figure \ref{fig-JSONiq-cleanup} shows how, in JSONiq, it is possible to open our previous example (in parallel), and structure every record to fit the schema, finally validating it to an output DataFrame. The output can either be written back to a structured format such as Parquet or Avro, or can be further used in more JSONiq code, for example, fed into ML pipelines also as shown in Figure \ref{fig-JSONiq-cleanup}.

RumbleDB was extended to support the JSound~\cite{JSound} schema language. JSound was designed to be formally precise, while simple, and supports object and array subtypes with clean semantics. JSound exists in a compact syntax for 90\% of the cases, and a more verbose syntax for the remaining ones. The compact syntax is designed to seamlessly map, in most cases, to a DataFrame schema. An example of schema compatible with DataFrames is shown in Figure \ref{fig-JSONiq-cleanup}, dynamically built as the second parameter of the annotate function. In other words, it is straightforward, when validating a sequence of items against this schema, for RumbleDB to store the resulting sequence in a DataFrame internally. Figure \ref{fig-data-type-mapping} shows the mapping between JSONiq types and Spark DataFrame schema types. RumbleDB was also extended to support the static definition of named, user-defined types that can also be used for static type analysis and further optimizations.

\begin{figure}[H]
\begin{tabularx}{\columnwidth}{|l|X|}
\hline
\textbf{JSONiq type} & \textbf{DataFrame type}   \\ \hline
byte                 & ByteType                                           \\ \hline
short                & ShortType                                           \\ \hline
int                  & IntegerType                                           \\ \hline
long                 & LongType                                           \\ \hline
boolean              & BooleanType                                           \\ \hline
double               & DoubleType                                            \\ \hline
float                & FloatType                                             \\ \hline
decimal              & DecimalType                                           \\ \hline
string               & StringType                                            \\ \hline
null                 & NullType                                              \\ \hline
date                 & DateType                                              \\ \hline
dateTime             & TimestampType                                         \\ \hline
date                 & DateType                                              \\ \hline
hexBinary            & BinaryType                                            \\ \hline
array (or subtypes)  & ArrayType                                             \\ \hline
object (or subtypes) & StructType                                            \\ \hline
\end{tabularx}
\caption{Mapping of JSONiq types to Spark DataFrame types upon validation.}
\label{fig-data-type-mapping}
\end{figure}

The generated SparkSQL DataFrame schema serves as the blueprint for mapping input Object Items into DataFrame rows. If the input sequence is physically an RDD or a DataFrame, the validation and mapping to a DataFrame are done in parallel. If it is a local (possibly non-materialized) sequence, then the DataFrame is newly created. 

\begin{figure}[!ht]
\centering
\begin{tikzpicture}[align=center, node distance=1mm and 6mm, line width=0.8pt] 
    \tikzstyle{free} = [inner sep=5pt]
    \tikzstyle{box} = [draw, rectangle, fill=myyellow, inner sep=5pt]
    
    \node[free] (dataset) {$\mathcal{D}$};
    \node[box] (tokenizer) [right=of dataset] {$\mathrm{tokenize}$};
    \node[box] (contain) [right=of tokenizer] {$\mathrm{contain}$};
    \node[box] (index) [above=of contain] {$\mathrm{index}$};
    \node[box] (convert) [right=of contain] {$\mathrm{convert}$};
    \node[free] (output) [right=of convert] {$\mathcal{D}_{clean}$};
    
    \draw[-stealth] (dataset) -- (tokenizer);
    \draw[-stealth] (tokenizer) -| ($(contain)-(1,0)$) |- (contain);
    \draw[-stealth] (tokenizer) -| ($(contain)-(1,0)$) |- (index);
    \draw[-stealth] (index) -| ($(contain)+(1,0)$) |- (convert);
    \draw[-stealth] (index) -| ($(contain)+(1,0)$) |- (convert);
    \draw[-stealth] (contain) -- (convert);
    \draw[-stealth] (convert) -- (output);
\end{tikzpicture}

\caption{A schematic depicting the data cleaning operations required to convert our raw data to $Clustered$ $YFCC$.}
\label{fig:processing}
\end{figure}
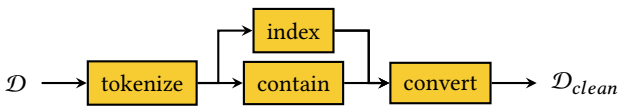

Finally, Figure \ref{fig-JSONiq-cleanup} shows how JSONiq also seamlessly handles purely textual data in order to convert a messy textual file (from the real world) and validate it to a DataFrame format.

Further examples of PySpark/spark.ml pipelines involving both data preparation and ML that would be also covered well by RumbleML can be found here \cite{Pafka2021}.

\section{The RumbleML data model}
\label{section-data-model}

The RumbleML framework aims to cover the full capabilities of the spark.ml library while providing a declarative API with JSONiq. The core idea of implementing RumbleDB as a highly expressive and performant facade to Spark is extended into the ML domain.

RumbleML's data model is based on (i) manipulating structured datasets as validated JSONiq sequences of objects, internally implemented as DataFrames; (ii) manipulating transformers and estimators to JSONiq function items, internally encapsulating spark.ml transformers and estimators; (iii) mapping the action of fitting and predicting to JSONiq dynamic function calls, internally forwarded to spark.ml method calls, with the output wrapped back into a JSONiq data model instance; and (iv) passing parameters to estimators and transformers, either at their creation or during the fitting/transforming phase, as JSONiq objects, internally converted to spark.ml ParamMaps.

\noindent\textbf{Transformers, estimators as function items}.
A central contribution of this paper is showing that the principal components of spark.ml, namely estimators and transformers, map seamlessly to JSONiq function items. 

Estimators and transformers are standardized according to the spark.ml pipeline API with specific methods for triggering execution. These methods are fit() for estimators and transform() for transformers. They share an identical signature for their arguments: The first argument is a DataFrame that serves as the input dataset, while the second is a spark.ml parameter map that contains a collection of ML parameters for tuning the operation. While their parameters are identical, return types of fit() and transform() methods are fundamentally different. A transformer is essentially a simple map function on an input dataset that returns an altered dataset. An estimator is, at the logical level, a higher-order function that returns a transformer, which is another function.

In the JSONiq data model, function items exhibit the standard nature of a function as they possess a name, parameters, a body to execute, and a return value. The executable body is the crux of function items as it can encapsulate any complex logic. For RumbleML, this executable body encapsulates estimator or transformer execution by calling, under the hood, the fit() or transform() method transparently to the user. Such a function item also accepts two parameters, which are forwarded to fit() or transform() method. Furthermore, a function item can have any sequence of items as its return type; for function items encapsulating estimators and transformers, the return type is more specifically a sequence of objects that, internally, is implemented as the DataFrame returned by spark.ml. This design harmoniously delegates computation to spark.ml under the hood, while exposing a functional and elegant API to RumbleDB users.

Figure \ref{fig-JSONiq-cleanup} shows code where both an estimator and transformer are created and then invoked on a training or test set.



\noindent\textbf{Creating estimators and transformers}.
Next, we explain how to create the function items that encapsulate estimators and transformers. In our first version of RumbleML, we chose to make it possible to look up an estimator or transformer by its name in spark.ml, in order to immediately support the entire spectrum of spark.ml features while being forward-compatible with future Spark versions. To this effect, RumbleML provides two static built-in functions get-estimator and get-transformer that can be seen in action in Figure \ref{fig-JSONiq-cleanup}.



In a future release of RumbleML, we consider increasing data independence by providing a library module with functions that provide a standard list of all estimators and transformers with names documented in RumbleML; these can then be mapped not only to spark.ml, but any other backend depending on its specific support, decoupling the code from the specific distributed ML backend used.

\noindent\textbf{Parameter maps as objects}.
In spark.ml the tuning parameters for estimators and transformers are exposed as a ParamMap object, which has a key-value format. This representation naturally maps to JSON objects. ParamMap objects exposed in strongly typed languages such as Java accommodate a variety of types. JSONiq being strongly typed, the types of almost all parameters exposed in spark.ml's parameter API map naturally to item types in the JDM, as shown in Figure \ref{tab:RumbleML_ParamMapping}. A few rare spark.ml parameters are not supported at the time of this paper. This decision is motivated by the fact that they appear in as little as 1\% of the collection of estimators and transformers offered by spark.ml.

\begin{figure}
\begin{tabularx}{\columnwidth}{|l|X|}
\hline
\textbf{Native Java Type} & \textbf{RumbleDB item type}                 \\ \hline
DataFrame                 & object*                                                           \\ \hline
ParamMap                  & object                                                            \\ \hline
boolean                   & boolean                                                           \\ \hline
double{[}{]}{[}{]}        & [ [ "double" ] ]                                                  \\ \hline
double{[}{]}              & [ "double" ]                                                      \\ \hline
double                    & double                                                            \\ \hline
float                     & double                                                            \\ \hline
int{[}{]}                 & [ "integer" ]                                                     \\ \hline
int                       & integer                                                           \\ \hline
long                      & double                                                            \\ \hline
String{[}{]}              & [ "string" ]                                                      \\ \hline
String                    & string                                                            \\ \hline
Matrix                    & [ [ "double" ] ]                                                  \\ \hline
Vector                    & [ "double" ]                                                      \\ \hline
Transformer               & function(object*, object) as object*                              \\ \hline
Estimator                 & function(object*, object) as function(object*, object) as object* \\ \hline

\end{tabularx}
\caption{spark.ml native Java parameter types and their corresponding item types in JSONiq. Array types are expressed with the JSound compact schema syntax, supported by RumbleDB, for readability.}
\label{tab:RumbleML_ParamMapping}
\end{figure}

\section{Implementation}
Let us first summarize the general architecture of RumbleDB. After parsing, translating, and code generation, the executable code has the form of a tree of runtime iterators. Coarsely, each runtime iterator corresponds to one logical JSONiq expression, although this correspondence is not exact for optimization purposes.

The runtime execution is triggered at the root iterator and recursively propagates to child iterators. At each level, RumbleDB switches seamlessly back and forth between all five execution modes presented in Section \ref{section-data-model}, based on the strategy decided during static analysis. The execution is lowered or lifted based on the wishes (via the API call) of the consuming iterator and the capabilities of the producing iterator.

\noindent\textbf{Static and dynamic function calls}.
RumbleML embraces functions for estimators and transformers. We thus extended RumbleDB with support for user-defined functions and function items, for static and dynamic function calls, and function name references. These features follow already established W3C standards, namely XQuery 3.0, and are a part of JSONiq.

An important challenge is how to select the execution modes for each static function call. Currently, we opt for the assignment, statically, of a specific execution mode to each function, recursively based on the statically known execution mode of each actual parameter and on the propagation of execution modes to the function body. In case of conflicts, a local execution mode takes precedence. Because of cycles due to JSONiq's support for recursive calls, this inference is done in multiple passes until it converges. For built-in functions, the execution mode is simply looked up in a catalog. 

For the ML use case, estimators and transformers are higher-order functions, and fitting or transforming involves a dynamic function call. In a dynamic function call, the function is computed dynamically from an expression of arbitrary complexity that resolves to a function item. This implies that the function to execute and its execution mode are not known until runtime. This is an undecidable problem, so we use a heuristic in our first prototype. In most cases, static type analysis will often be able to determine the signature of the function item being called. If it determines that the return type of this function item has an arity of one, then the execution mode is local. This happens in particular when the function item is an estimator, as it returns the trained model as a single function item. Otherwise, it may be a transformer and thus, if the first parameter supports DataFrame execution mode, then the dynamic function call does as well: we assume it returns the DataFrame enriched with predictions. Else, its execution is local. If the compile-time assumption is found to be incorrect at runtime, a dynamic error is thrown. We found that this heuristic worked in all our use cases, and will improve it if necessary in future iterations.

\noindent\textbf{Estimator and transformer lookup}.
RumbleML makes builtin estimators and transformers available via two builtin functions "get-estimator()" and "get-transformer()". Both take two parameters: the name of the desired estimator or transformer, and an object containing parameters. For an estimator, the return type of the generated function item is a function type. For a transformer, its return type is a sequence of objects. In either case, these are singleton items, so the execution mode is the single-item mode.

RumbleML interprets the first string parameter as the simple spark.ml class name of the estimator or transformer; it resolves it to its fully qualified name, which it uses to instantiate it as a Java object with the Java Reflection API. RumbleML then sets its parameters with the parameter object supplied as the second parameter of the lookup function. Parameters are explained in Section \ref{section-parameters}

\noindent\textbf{Estimator and transformer execution}.
\label{sec:estimator_execution}
The body of the function returned by the lookup is a runtime iterator encapsulating the spark.ml estimator or transformer. When the function item is invoked dynamically, the iterator forwards the input to this estimator (fit) or transformer object (transform), letting spark.ml do the computations and map the output back to a JSONiq value. Section \ref{section-parameters} explains how the second parameter is processed An exception is thrown if the sequence of objects is not physically a DataFrame. We explained in Section \ref{section-data-cleaning} how users can validate datasets with a JSound schema in order to prepare their data as a DataFrame.

In the case of an estimator, the resulting model, a transformer, is encapsulated inside a newly generated function item in a similar way to the lookup of built-in transformers described above. Since the end result is a single item as determined at compile-time, it is executed with the single-item execution mode. In the case of a transformer, the result is a transformed DataFrame, exposed as a sequence of objects, meaning that the execution mode of this iterator is DataFrame-based as determined at compile time.

\noindent\textbf{Feature vectors}.
spark.ml enriches Spark's DataFrame type system with dense and sparse vectors to store features. The user creates new feature columns in these formats by invoking the VectorAssembler transformer with the name of the additional column, like any other transformer. Then, this column name is passed as a parameter to the next transformer or estimator in the pipeline. This is a very natural workflow for the end ML user, and we also used this in our experiments. On the logical level, these types are exposed to the RumbleML user as JSONiq arrays of doubles, materialized only if necessary.

\noindent\textbf{Estimator and transformer parameters}.
\label{section-parameters}
The parameter object passed as the second argument to many functions, allows the user to specify parameters that modify the behavior of fitting or transforming. It is converted by RumbleML to a spark.ml ParamMap that is passed to the encapsulated spark.ml estimator or transformer. spark.ml's ParamMap API is standardized to be shared by all the estimators and transformers.

RumbleML contains a mapping from each kind of estimator and transformer to its parameters and the types thereof and validates that the supplied parameters are valid before setting them, i.e., that they apply to the estimator or transformer, and that their types are correct. If the parameters are valid, then each argument value provided as a JSONiq item is converted to the correct Java type.

\noindent\textbf{Optimizations}.

In order to match the performance of an ML pipeline using spark.ml with an imperative language, several optimizations were necessary.

One of them was described earlier and is the static choice of execution mode to detect fitting and transforming to preserve the use of DataFrames. Indeed, in a previous round of iteration, all dynamic function calls were executed locally. This forced a materialization of the DataFrame output by a transformer, which either led to an error if its size exceed the materialization cap, or to a slower, local execution in any code consuming this DataFrame, or to extra code to re-validate and re-create a DataFrame for use in subsequent pipelines. The static detection of execution calls solves all three issues in the majority of the cases in which the static detection succeeds.

Another optimization was to extend the implementation of some language features, such as predicates or FLWOR expressions, so that they can produce a DataFrame upon DataFrame input, using Spark SQL either with UDFs or even natively. For example, the optimization of a predicate is always achievable with at least a User-Defined Function encapsulating the predicate expression, and in some cases can even directly be mapped to native Spark SQL, making it even faster.

A study of the impact of both optimizations on the performance of RumbleML is provided in Section \ref{section-experiments}.

\section{Measurements and results}
\label{section-experiments}
\subsection{Experimental Setup} 

\noindent\textbf{Platform.}~~%
We conduct all experiments on a cluster of three m5.4xlarge instances in AWS with EMR 6.4.0 and Spark 3.1.2 installed. These instances run on machines with Intel Xeon® Platinum 8175 3.1GHz CPUs and have 8 GiB of main memory per
CPU core, and, for the instance sizes we use, get up to 10 Gbps of
network bandwidth. Each instance has eight physical CPU cores and unless otherwise mentioned, we process the data directly off of S3.

\noindent\textbf{Datasets.}~~%
We train and test our experiments on two large-scale datasets. ${Criteo}$~\cite{lerallut2015large} is a sparse dataset of 50 GB, that contains feature values that represent click feedback from millions of display ads. The data set contains 98M tuples with binary labels that classify if a user has clicked on an ad or not. We use 92M tuples for training and 6M for evaluation. Next, we use $Clustered$ $YFCC$, a dense dataset of 50 GB, that contains 1M tuples of feature vector representations (4096-dim) extracted from images of the YFCC100M dataset~\cite{thomee2016yfcc100m}. For our experiments, we split the $Clustered$ $YFCC$ data set equally into train and test data. The test set is chosen to be 25 GB in order to evaluate the impact of large inference and score calculation at test time.

\noindent\textbf{Data cleaning.}~~%
To create $Clustered$ $YFCC$, several data cleaning operations are required to create a standardized parquet format. The raw data initially concatenates fc7 vectors of a pre-trained CNN together with its multi-class predictions (see Figure \ref{fig-messy-dataset}). Thus, our data cleaning process first tokenizes and splits the text into predictions and features. Second, we check if predictions contain the label $outdoor$ or $indoor$ and convert it into a binary label. Third, we index each feature to create a libsvm format. Lastly, we concatenate the output and compress the result into a parquet file (see Figure \ref{fig:processing}).

\noindent\textbf{Models.}~~%
We train and evaluate a Logistic Regression, Linear SVC, RandomForest, and Naive Bayes on $Criteo$ and $Clustered$ $YFCC$. All experimental results are run $5$ times and aggregated together.

\begin{figure}[!ht]
\centering
\begin{tikzpicture}[align=center, node distance=5mm and 10mm, line width=0.8pt] 
    \tikzstyle{free} = [inner sep=5pt]
    \tikzstyle{box} = [draw, rectangle, fill=myblue, inner sep=5pt]
    
    \node[free] (dataset) {$\mathcal{D}$};
    \node[box] (scaler) [right=of dataset] {$\mathrm{scaler}$};
    \node[box] (model) [right=of scaler] {$\mathrm{model}$};
    \node[box] (scorer) [right=of model] {$\mathrm{scorer}$};
    \node[free] (acc) [right=of scorer] {$u$};
    
    \draw[-stealth] (dataset) -- (scaler);
    \draw[-stealth] (scaler) -- (model);
    \draw[-stealth] (model) -- (scorer);
    \draw[-stealth] (scorer) -- (acc);
\end{tikzpicture}

\caption{A schematic depicting the operations that make the scaler-model pipeline.}
\label{fig:pipe-1}
\end{figure}
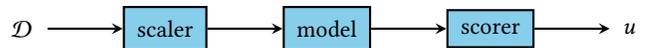

\noindent\textbf{Pipelines.}~~%
We randomly subsampled a real-world dataset of 500K ML.NET pipelines and converted 12 end-to-end pipelines (see Figure \ref{fig:pipe-1}) into JSONiq and spark.ml \cite{karlas2020vldb}. The first six pipelines are model pipelines trained on a sparse dataset ($Criteo$) without (P1-P3) and with an absolute scaler operator (P4-P6). The next six pipelines are trained on a a dense dataset ($Clustered$ $YFCC$) without (P7-9) and with an absolute scaler operator (P10-P12). 

\subsection{Code comparison}
The YFCC pipeline script shown in Figure \ref{fig-JSONiq-cleanup} was also written in PySpark and spark.ml in order to compare the languages with a few metrics. In practice, we insist that, because of the complexity that it involves, data scientists use different tools rather than a single script.

We then compared the two scripts using different metrics as shown in Figure \ref{tbl:requirements-summary}. Thus, beyond considerations of functional vs. imperative, it can be seen that JSONiq exposes more compact and expressive constructs than PySpark and spark.ml.

\begin{figure}
  \centering%
  \newcommand{\system}[1]{\rotatebox[origin=c]{0}{\textbf{#1}}}%
  \newcommand{\w}{\textasteriskcentered}%
  \newcommand{\ww}{\textasteriskcentered\textasteriskcentered}%
  \newcommand{\www}{\textasteriskcentered\textasteriskcentered\textasteriskcentered}%
  \resizebox{.6\linewidth}{!}{%
  \begin{tabular}{@{}l@{\hspace*{0pt}}ccccccc@{}}
    \toprule
      & \system{JSONiq} & \system{SparkML} \\
    \midrule
    \#characters                        & \textbf{857}   & 1146 	\\
    \#lines                             & \textbf{31}    & 39          	\\
    \#clauses                           & \textbf{37}   & 66          	\\
    \#unique clauses                    & \textbf{22}    & 26           	\\
    \bottomrule				
  \end{tabular}%
  }%
  \caption{Summary of pipeline scripts for RumbleML and spark.ml.}%
  \label{tbl:requirements-summary}%
\end{figure}

\subsection{Performance comparison}
\begin{figure}[H]
\centering
\includegraphics[width=\columnwidth]{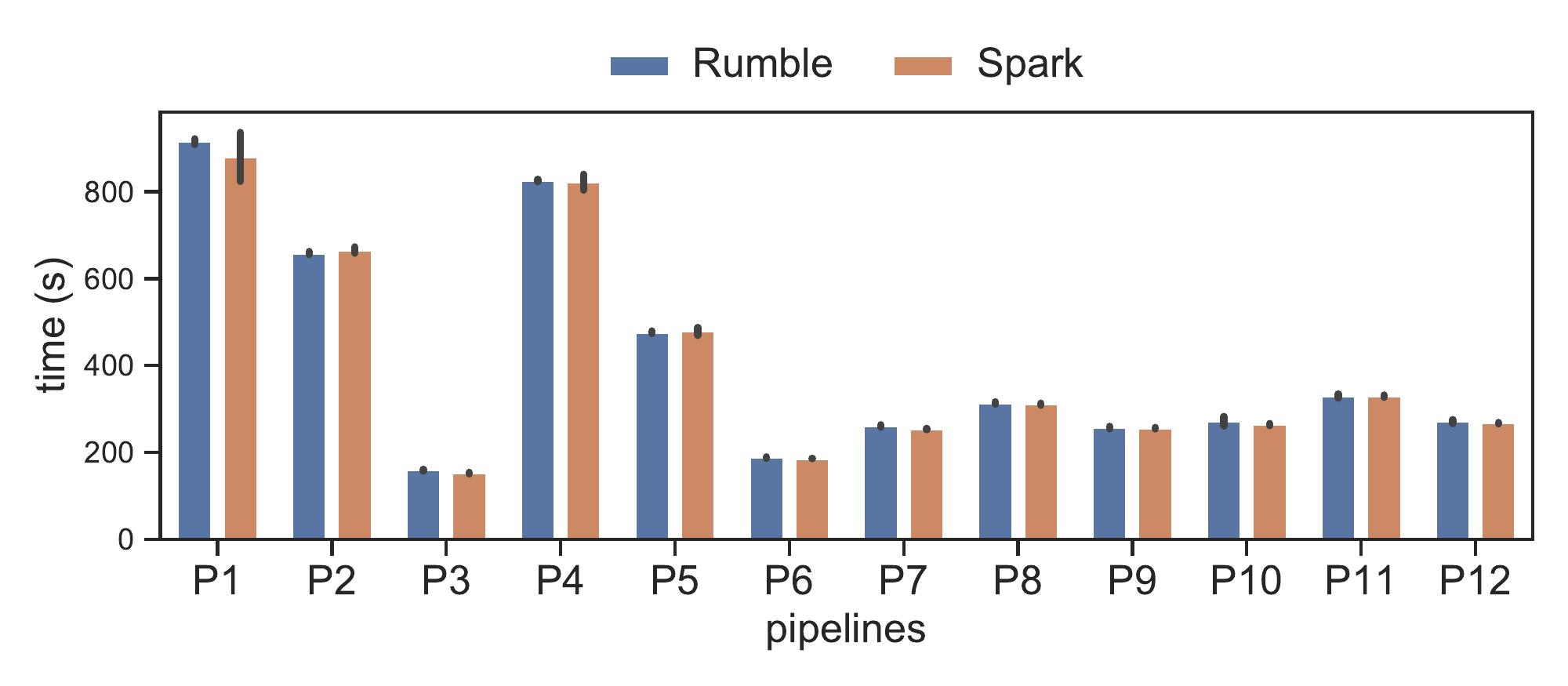}
\caption{End-to-End performance on ML pipelines}
\label{fig:all}
\end{figure}

In the following, we report the runtime performance and accuracy of our experiments described above.
Note that for both, Spark and Rumble, Random Forest did not converge for the sparse $Criteo$ dataset with our initial cluster setup and parameters chosen. We believe this is due to the underlying tree optimizer that cannot cope with 1M sparse feature values. 
Furthermore, Naive Bayes models did not support pipelines trained on the $Clustered$ $YFCC$ as it contains negative feature values. Figure \ref{fig:all} show the end-to-end runtime for $Criteo$ and $Clustered$ $YFCC$ respectively. We did not observe any runtime differences between spark.ml and RumbleML that can't be explained by variance. Furthermore, we did not observe any differences in the resulting accuracy.

\subsection{Impact of Optimizations}
\begin{figure}[H]
\centering
\includegraphics[width=\columnwidth]{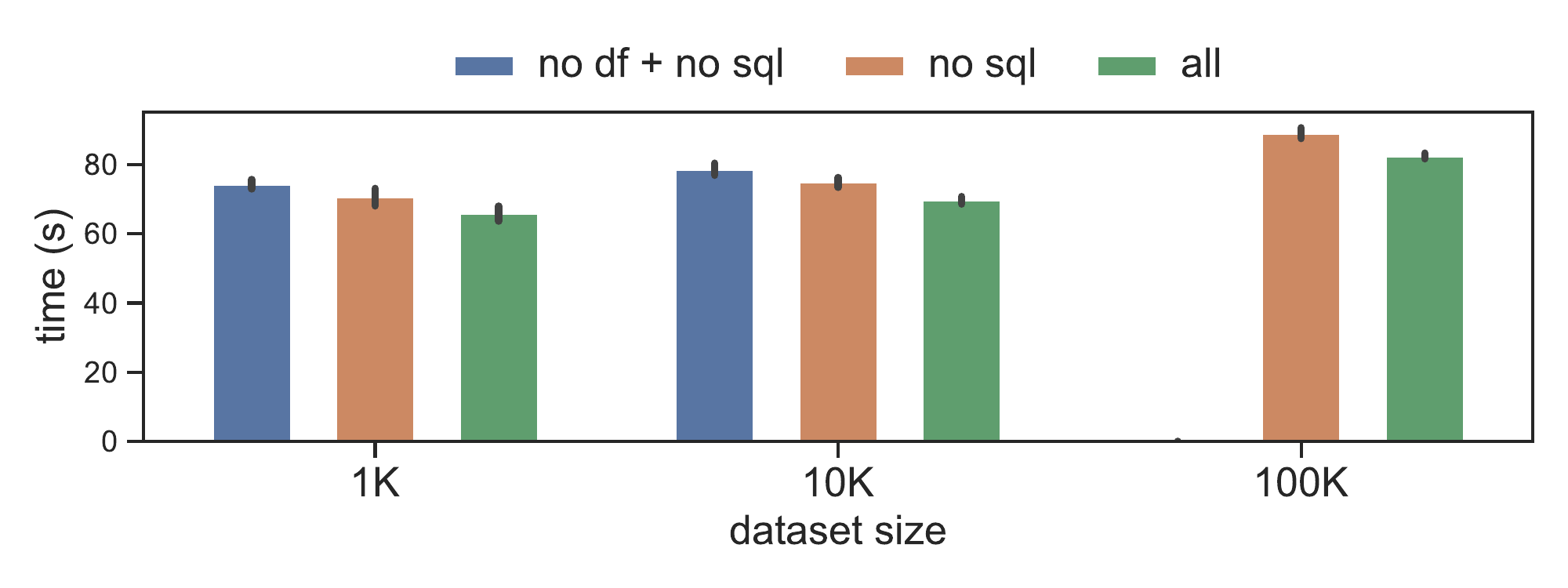}
\caption{Ablation study on Criteo (P1)}
\label{fig:ablation_study}
\end{figure}

In the following, we report runtime performance for RumbleML with and without optimizations on the Criteo dataset (see Figure \ref{fig:ablation_study}). Without any optimization (blue), RumbleML can only execute queries locally, resulting in loss of performance and critical failure due to memory limitations when the data size increases to larger than 10K data points (thus no time measurements were able to be performed for 100K). With DataFrame optimization turned on, RumbleML is able to leverage the cluster structure (orange). Finally, by integrating optimizations for filter queries (green), RumbleML achieves a performance that is comparable to native spark.ml.

\section{Related and future work}

Another high-level language for ML and AI use cases is the inkling language, part of the Bonsai platform ~~\cite{Inkling}. Alternative languages for the cleanup of denormalized data include XQuery ~\cite{boag2002xquery}, AQL~\cite{grahne1998aql}, SQL++~\cite{ong2014sql++}, PartiQL~\cite{PartiQL}, GraphQL~\cite{hartig2018semantics}, however most of them do not support higher-order functions out of the box. There are also other platforms than spark.ml, for example, TensorFlow ~\cite{abadi2016tensorflow}, PyTorch~\cite{paszke2017automatic} and Weka~\cite{hall2009weka}, although, from our perspective, our system can also be extended to build a JSONiq layer on top of them. Also, none of the above systems cover both data preparation and ML pipelines with a single, integrated functional language.

In future versions of RumbleML, we plan to extend it with the ability for users to program their own ML algorithms directly in JSONiq, that is, to implement estimators and transformers as regular user-defined functions in JSONiq with the same signatures. Syntactically, this is already possible, and the work will mostly consist in optimizing RumbleDB to automatically detect them. We will also consider extending the support of transformers to take more generic sequences as input (e.g., sequences of strings), in order to integrate the data cleaning stage even more seamlessly into the syntax to build the pipeline. We also plan to enhance static schema detection in order to make user-defined schemas superfluous in many cases.

We also plan to offer complete coverage of spark.ml's estimators, transformers, and parameters by extending the mapping to the rarer types.

\section{Conclusion}
RumbleML is available as a free, open-source product, currently in beta. The successful implementation of RumbleML proves yet again that the JSONiq language is a good fit for programming lakehouse systems and that it can be implemented with no loss of performance, but gaining in productivity.

\addtolength{\textheight}{-4cm}

\printbibliography




\end{document}